\documentclass[twocolumn,showpacs,preprintnumbers,amsmath,amssymb]{revtex4}

\usepackage{graphicx}
\usepackage{dcolumn}
\usepackage{bm}

\begin{document}

\preprint{APS/123-QED}

\title{$^{59}$Co and $^{75}$As NMR Investigation of Lightly Doped Ba(Fe$_{1-x}$Co$_x$)$_2$As$_2$ (x =0.02, 0.04)}

\author{F. L. Ning$^{1}$, K. Ahilan$^{1}$, T. Imai$^{1,2}$, A. S. Sefat$^{3}$, R. Jin$^{3}$, M.A. McGuire$^{3}$, B.C. Sales$^{3}$, and D. Mandrus$^{3}$}

\affiliation{$^{1}$Department of Physics and Astronomy, McMaster University, Hamilton, Ontario L8S4M1, Canada}
\affiliation{$^{2}$Canadian Institute for Advanced Research, Toronto, Ontario M5G1Z8, Canada}
\affiliation{$^{3}$Materials Science and Technology Division, Oak Ridge National Laboratory, TN 37831, USA}

\date{\today}

\begin{abstract}
We investigate the electronic properties of
Ba(Fe$_{1-x}$Co$_x$)$_2$As$_2$ (x = 0.02, 0.04) in the lightly
electron-doped regime by $^{59}$Co and $^{75}$As NMR. We demonstrate
that Co doping significantly suppresses the magnetic ordering
temperature to the SDW state, T$_{SDW}$. Furthermore, ordered
moments below T$_{SDW}$ exhibit large distribution. Strong spin
fluctuations remain even below T$_{SDW}$, persisting all the way
down to 4.2 K. We find no signature of additional freezing of spin
degrees of freedom unlike the case of the lightly hole-doped stripe
phase of the cuprates.
\end{abstract}


\maketitle


The recent discovery of high transition temperature (high-T$_c$)
superconductivity in iron-pnictides \cite{Kamihara, RenZA, Rotter,
Athena1} has spurred huge excitement in the condensed matter physics
community. The FeAs layers, consisting of a square-lattice of Fe
coordinated by four As, are the crucial component responsible for
the superconductivity. The quasi-two dimensional layered structure
is reminiscent of the CuO$_2$ layers in the high-T$_c$ cuprates, but
many dissimilarities exist between iron-pnictides and cuprates. For
example, doping $\sim$ 4 $\%$ of impurities into cuprates could
destroys their superconductivity \cite{Fukuzumi}, but doping Co or
Ni into the FeAs layers of BaFe$_2$As$_2$ $\textit{induces}$
superconductivity with T$_c$ as high as 22 K \cite{Athena1,XuZA}.
Earlier studies showed that a prototypical parent compound of
iron-pnictides, BaFe$_2$As$_2$ (x = 0) is an itinerant
antiferromagnet, and exhibits simultaneous first order structural
and magnetic phase transitions at T$_{SDW}$ $\sim$ 135 K
\cite{Rotter2,HuangQ, Takigawa, Iyo}. 2$\%$ and 4$\%$ Co doping into
BaFe$_2$As$_2$ quickly suppresses the ordering temperatures to
T$_{SDW}$ $\sim$ 100 K and 66 K respectively \cite{Ahilan, Ning2,
Canfield, Fisher, ChenXH}. When the doping level is increased to
$\sim$ 8 $\%$, superconductivity appears with optimized T$_c$ $\sim$
22 K. Very little is known about the nature of the magnetically
ordered state below T$_{SDW}$ in the presence of 2 - 4 $\%$ electron
doping.

In this Rapid Communication, we will report a microscopic
investigation by Nuclear Magnetic Resonance (NMR) on the electronic
properties of lightly electron-doped Ba(Fe$_{1-x}$Co$_x$)$_2$As$_2$
(x = 0.02, 0.04). We will show that Co doping suppresses the
magnetic ordering temperature, T$_{SDW}$. Furthermore, as little as
2 $\%$ Co doping transforms the nature of the ground state from the
Commensurate Spin Density Wave (C-SDW) state observed in the undoped
parent compound BaFe$_2$As$_2$ \cite{Rotter2,HuangQ,Takigawa} to a
different state, most likely a highly disordered Incommensurate Spin
Density Wave (IC-SDW) state. We will show that strong spin
fluctuations remain below T$_{SDW}$ all the way down to 4.2 K. There
is no signature of additional freezing of spin degrees of freedom in
contrast with the case of the lightly doped stripe phase of the
cuprates \cite{Chou, Cho}.

We grew the single crystals with x = 0, 0.02 and 0.04 from FeAs flux
\cite{Athena1} and determined the actual Co concentration by EDS
(Energy Dispersive X-Ray Spectroscopy). These are the identical
pieces that were used for our previous $^{75}$As NMR study in the
paramagnetic state \cite{Ning2,Ning1}. We carried out NMR
measurements using the standard pulsed NMR techniques on either one
piece of crystal (x = 0, 0.04) or aligned crystals (x = 0.02, two
pieces) with total masses of $\sim$ 2 to $\sim$ 20 mg.
\begin{figure}[!htpb]\vspace{-0.5cm}
\includegraphics[width = 10cm, angle =0]{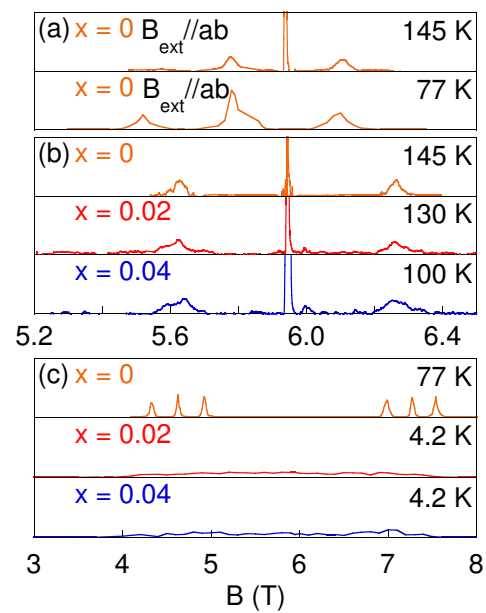}\vspace{-4.5cm}
\caption{(Color Online). $^{75}$As field swept NMR lineshapes of
Ba(Fe$_{1-x}$Co$_x$)$_2$As$_2$ measured at \textit{f} = 43.503 MHz,
for x = 0 (T$_{SDW}$  = 135 K), x = 0.02 (T$_{SDW}$ = 100 K) and x =
0.04 (T$_{SDW}$ = 66 K). B$_{ext}$ was applied along the
\textit{c}-axis, except in panel (a) where B$_{ext}$//ab. Notice
that the positions of the NMR lines in the paramagnetic state only
shift from 145 K to 77 K in (a) because the hyperfine magnetic field
is along the \textit{c}-axis. In (c), NMR lines either split(x =0)
or broaden (x = 0.02, 0.04).} \label{fig:Ning_fig1:epsart}
\end{figure}

\begin{figure}[!htpb]
\includegraphics[width = 10 cm, angle =0]{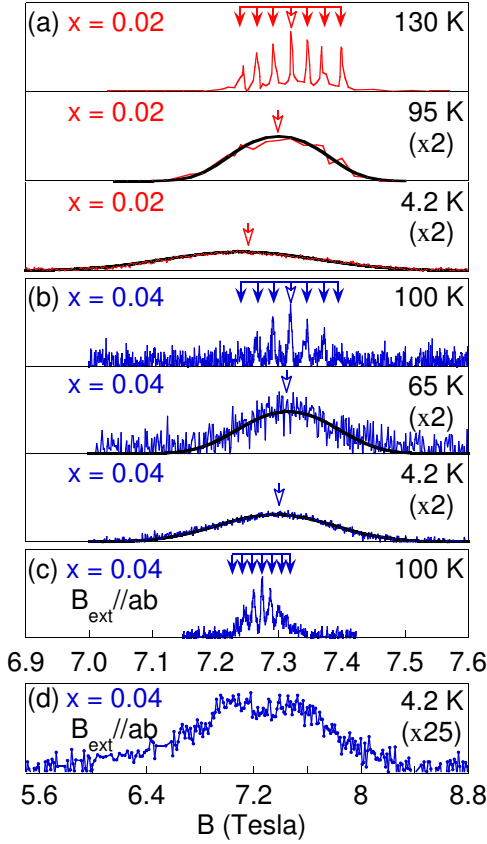}\vspace*{-1.5cm}
\caption{\label{fig2:epsart} (Color Online). $^{59}$Co field swept
NMR lineshapes of Ba(Fe$_{1-x}$Co$_x$)$_2$As$_2$ measured at
\textit{f} = 74.103 MHz for (a) x = 0.02 at 130 K (paramagnetic), 95
K and 4.2 K (SDW); (b) x = 0.04 at 100 K (paramagnetic), 65 K and
4.2 K(SDW). The overall intensities in the SDW state have been
amplified by a factor of 2 compared with those in the paramagnetic
state. Solid curves represent Gaussian fit to the data. In (c) and
(d), B$_{ext}$ was applied within the \textit{ab}-plane, otherwise
B$_{ext}$ was applied along the \textit{c}-axis. Notice the scale of
the horizontal axis in panel (d) is expanded. The overall intensity
in (d) has been amplified by 25 times compared with (c). Open arrows
mark where we measured $T_1$.}
\end{figure}

In Fig. 1, we present the typical field swept lineshapes at a fixed
frequency $\textit{f}$ = 43.503 MHz for $^{75}$As (nuclear spin I =
$\frac{3}{2}$, $\gamma_n/2\pi$ = 7.2919 MHz/Tesla) with external
field B$_{ext}$ applied either along the $\textit{c}$-axis
(B$_{ext}$//c) or within the $\textit{ab}$-plane (B$_{ext}$//ab).
The nuclear spin Hamiltonian can be expressed as a summation of the
Zeeman and nuclear quadrupole interaction terms,\begin{equation}
\label{1} \ \textit{H} = -\gamma_n \textit{h}\textbf{B} \cdot
\textbf{I} + \frac{\textit{h} \nu_Q^c}{6}\{3I_z^2- I(I+1) +
\frac{1}{2}\eta(I_+^2+I_-^2)\},
\end{equation}where $\textit{h}$ is Planck's constant, and $\textit{I}$ is the
nuclear spin. $\textbf{B}$ is the local field at the observed
nuclear spin, and the summation of the external field
$\textbf{B$_{ext}$}$ and the hyperfine field $\textbf{B$_{hf}$}$
from the ordered moments. The nuclear quadrupole interaction
$\nu_Q^c$ is proportional to the Electric Field Gradient (EFG) at
the observed As site, and $\eta$ is the asymmetry parameter of the
EFG, $\eta$ = $|\nu_Q^a-\nu_Q^b|/\nu_Q^c$.

First, we briefly discuss the $^{75}$As NMR results in undoped
BaFe$_2$As$_2$ ( x = 0, T$_{SDW}$ = 135 K). Each $^{75}$As site
gives rise to three transitions from I$_z$ = $\frac{2m+1}{2}$ to
$\frac{2m-1}{2}$ (where m = -1, 0, 1) in the paramagnetic state, as
shown in Fig. 1(a) and (b). The fact that we observe only one set of
$^{75}$As NMR signals above T$_{SDW}$ is evidence that there is only
one type of As site in the undoped parent compound. The satellite
transitions (m = -1, 1) are somewhat broader than the central peak
(m = 0), but are still fairly sharp, implying that $\nu_Q^c$ has a
well defined value. From the split between the main peak and the
satellite peaks in Fig. 1(a), $\Delta$B $\sim$ 0.162 Tesla, we
estimate $^{75}\nu_Q^{ab}$ = $^{75}\gamma_n$$\Delta$B = 1.188 MHz.
From the split in Fig. 1(b), we estimate $^{75}\nu_Q^c$ = 2.3 MHz.
Thus $\eta$ $\cong$ 0 at 145 K for $^{75}$As, as expected for the
tetragonal symmetry. At 77 K, the $^{75}$As NMR lines with
B$_{ext}$//c split into two sets as shown in Fig. 1(c). This is
because the hyperfine field at $^{75}$As site from the ordered
moments, $\pm$ B$_{hf}$, is along the $\textit{c}$-axis. For
B$_{ext}$//ab, the $^{75}$As line only shifts to the lower field
side, because the resonance condition is satisfied as
$^{75}\gamma_n$$\sqrt{B_{ext}^2 + B_{hf}^2} = \textit{f}$. These
results confirm that the hyperfine field on the As site is along the
$\textit{c}$-axis \cite{Takigawa,Iyo}. The relatively sharp peaks at
77 K in the ordered state indicate that the ordered moments are
commensurate with the lattice and the hyperfine field has only two
discrete values, e.g. B$_{hf}^c$ = $\pm$1.32 Tesla at 77 K
\cite{Takigawa}.

In Fig. 1(b), we also show the influence of Co doping in the
paramagnetic state above T$_{SDW}$. The lineshape for the doped
samples are very similar to the undoped case, except that the
satellite transitions become broader due to additional distribution
of $^{75}\nu_Q^{c}$ caused by the disorder in the lattice
environment. The magnitude of $^{75}\nu_Q^{c} \sim 2.3$~ MHz is by a
factor of $\sim 5$ smaller than the case of LaFeAsO$_{1-\delta}$
\cite{Mukuda}. This is presumably because $^{75}$As ions  are
surrounded by 2+ ions only (Fe$^{2+}$ and Ba$^{2+}$) in the present
case, while $^{75}$As ions in LaFeAsO$_{1-\delta}$ have $La^{3+}$
and O$^{2-}$ ions nearby, in addition to Fe$^{2+}$ ions; the charge
disparity would enhance the EFG, hence $^{75}\nu_Q^{c}$ in
LaFeAsO$_{1-\delta}$. We also note that$^{75}\nu_Q^{c} \sim 2.3$~
MHz is nearly independent of the level of doping, and there is no
evidence for correlation between $^{75}\nu_Q^c$ and $T_c$. This is
in contrast with the case of in LaFeAsO$_{1-\delta}$ where $T_c$
appears to have a strong correlation with the $^{75}\nu_Q^{c}$
\cite{Mukuda}. On the other hand, we found that the lineshapes are
qualitatively different between undoped and doped samples below
T$_{SDW}$, as shown in Fig. 1(c). Unexpectedly, the $^{75}$As lines
do not split in 2 \% and 4 \% Co doped samples. Instead, the
$^{75}$As NMR lines broaden, and become almost featureless. The spin
echo signal could be detected everywhere between 4 and 7.5 Tesla,
which implies that $|B_{hf}^c|$ at $^{75}$As sites is continuously
distributed from 0 to $\lesssim$ 1.32 Tesla.

In Fig. 2, we present the typical field swept $^{59}$Co (nuclear
spin I = $\frac{7}{2}$, $\gamma_n/2\pi$ = 10.054 MHz/Tesla)
lineshapes with B$_{ext}$//c or B$_{ext}$//ab. Co is randomly doped
into FeAs layers by replacing Fe. The probability for each Co to
have four Fe at the nearest neighbor (n.n.) sites is 92.2 $\%$ for x
= 0.02 and 84.9 $\%$ for x = 0.04, respectively. Thus the Co NMR
lineshape is dominated by the NMR signals from the Co with four n.n.
Fe, and the Co NMR line splits into seven peaks separated by
$^{59}\nu_Q^{c}$. We estimate $^{59}\nu_Q^c$ $\sim$ 0.26 MHz,
$^{59}\nu_Q^{ab}$ $\sim$ 0.13 MHz and $\eta$ = 0 for both x =0.02
and 0.04.
\begin{figure}[!htpb]\vspace{-1.5cm}
\centering\includegraphics[width = 10cm, angle =0]{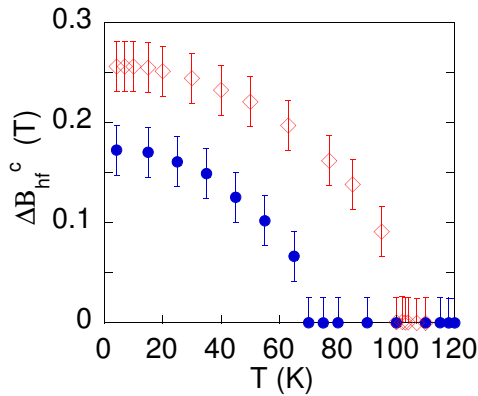}\vspace{-1.5cm}
 \caption{ (Color Online). The temperature dependence of the magnetic broadening $\Delta$B$_{hf}^c$ of $^{59}$Co NMR lines for x = 0.02 ($\diamond$),
x = 0.04($\bullet$).} \label{fig:Ning_fig1:epsart}
\end{figure}
\begin{figure}[!htpb]\vspace{-1cm}
\includegraphics[width = 10cm, angle =0]{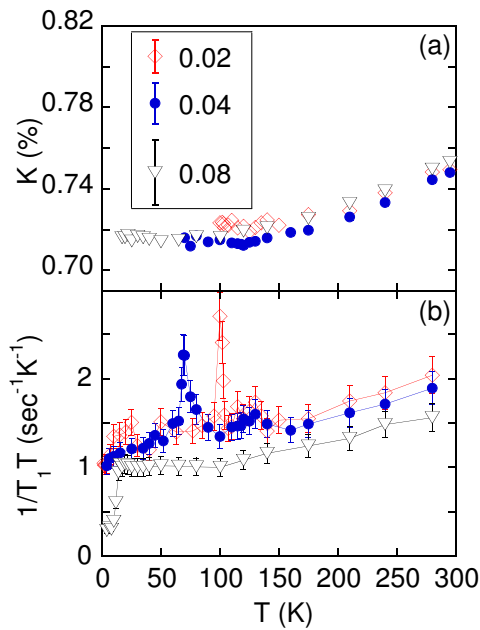}\vspace{-4.5cm}
\caption{ (Color Online). (a) $^{59}$Co Knight Shift for x = 0.02
($\diamond$), 0.04 ($\bullet$) and 0.08 ($\triangledown$). The error
bars in the legend represent the FWHM (Full Width at Half Maximum)
of Co center transition at 295 K. (b) $\frac{1}{T_1T}$ at $^{59}$Co
sites for x = 0.02 ($\diamond$), 0.04 ($\bullet$) and 0.08
($\triangledown$). We measured $T_1$ at the central transition, as
shown by the open arrows in Fig. 2.} \label{fig:Ning_fig1:epsart}
\end{figure}

Below T$_{SDW}$, the $^{59}$Co NMR lines become broader and the
seven discrete peaks caused by the quadrupole split $^{59}\nu_Q^c$
are smeared out. The whole NMR line becomes completely featureless
at low temperatures. We observe no signature of residual sharp peaks
below T$_{SDW}$, hence all $^{59}$Co nuclear spins are under the
influence of magnetic ordering. The integrated intensity corrected
for the Boltzmann factor agree well between 4.2 K and 100 K, hence
we observe all $^{59}$Co nuclear spins at 4.2 K. This conservation
of the total intensity rules out any possibility of phase separation
or macroscopic inhomogeneity in the sample. Close inspection of the
line positions reveals that the center of the broad line
progressively shifts to the lower field side with decreasing
temperature when we apply B$_{ext}$ along the $\textit{c}$-axis. For
example, the center transition for x = 0.04 at 100 K is at B$_{ext}$
$\sim$ 7.318 Tesla, which shifts by 0.023 Tesla to 7.295 Tesla at
4.2 K. On the other hand, the $^{59}$Co NMR lines for x = 0.04 split
into two broad humps when B$_{ext}$ is applied along the
$\textit{ab}$-plane instead, as shown in Fig. 2(d). The separation
between the center of the two broad humps, $\sim$ 0.6 Tesla, is much
larger than the small shift, 0.023 Tesla, observed along
B$_{ext}$//c . This implies that the hyperfine field at the
$^{59}$Co site is primarily within the $\textit{ab}$-plane. Combined
with the fact that $^{75}$As lines do not exhibit splitting with
B$_{ext}$//c, we conclude that Co doping changes the C-SDW spin
structure of BaFe$_2$As$_2$.

We fit the broad, featureless $^{59}$Co NMR lineshapes with
B$_{ext}$//c by assuming that the quadrupole splitting by
$^{59}\nu_Q^c$ does not depend on temperature and all seven
transitions become broader by a Gaussian distribution of the
hyperfine fields $\Delta$B$_{hf}^c$ below T$_{SDW}$. The fits are
reasonable for both x = 0.02 and 0.04, and we were able to deduce
the Gaussian width $\Delta$B$_{hf}^c$ as summarized in Fig. 3.
$\Delta$B$_{hf}^c$ continuously increases and finally saturates at
base temperature.

In Fig. 4(a), we present the temperature dependence of the static
spin susceptibility, $\chi$, for x = 0.02 and 0.04 as measured by
$^{59}$Co NMR Knight shift, K. We also plot the result for the
superconducting x = 0.08 sample for comparison \cite{Ning1}. In
general, we can write K = K$_{spin}$ + K$_{chem}$. K$_{spin}$ is the
spin contribution, which is proportional to the local spin
susceptibility $\chi$, while K$_{chem}$ is the
temperature-independent chemical shift. K$_{chem}$ is not related to
$\chi$. Our results indicate that $\chi$ gradually decreases below
$\sim$ 300 K, and begins to level off below $\sim$ 100 K. This is
consistent with our earlier results based on $^{75}$As NMR
\cite{Ning2}. The $^{59}$Co NMR linewidth is too broad to determine
the concentration dependence accurately.

In Fig. 4(b), we show the temperature dependence of $\bf{q}$
integrated dynamical spin susceptibility as measured by
$\frac{1}{T_{1}T} \propto \sum_{{\bf
q}}{|A_{hf}(\bf{q})|^{2}\frac{\chi"({\bf q}, \textit{f
})}{\textit{f}}}$ at $^{59}$Co sites, where $|A_{hf}(\bf{q})|^{2}$
is the wave-vector $\bf{q}$-dependent hyperfine form factor
\cite{Ning1}, $\chi"$(${\bf q},f)$ is the imaginary part of the
dynamical electron spin susceptibility (i.e. spin fluctuations), and
$\textit{f}$ is the NMR frequency ($\lesssim$ 10$^2$ MHz).
$\frac{1}{T_1T}$ shows a divergent behavior at $\sim$ 100 K for x =
0.02, and $\sim$ 66 K for x = 0.04. These temperatures agree well
with the maximum negative slope observed for in-plane resistivity
\cite{Ning2}. In ref [12], we also reported the divergent behavior
of $^{75}(\frac{1}{T_1T})$ at $^{75}$As sites with
B$_{ext}$//c-axis. In this geometry, $^{75}(\frac{1}{T_1T})$ probes
spin fluctuations within the $\textit{ab}$-plane. On the other hand,
Kitagawa et al \cite{Takigawa} showed that the $^{75}$As hyperfine
form factor satisfies $|^{75}A_{hf}(\bf{q})|^{2}$ = 0 within the
$\textit{ab}$-plane for commensurate spin fluctuations due to
cancelation of the transferred hyperfine fields. Therefore, these
$\frac{1}{T_1T}$ data at $^{59}$Co and $^{75}$As provide strong
evidence for the critical slowing down of the incommensurate spin
fluctuations toward a second order phase transition at T$_{SDW}$.
Interestingly, $\frac{1}{T_1T}$ decreases roughly linearly with
temperature down to the base temperature for both x = 0.02 and 0.04
except near T$_{SDW}$, and shows qualitatively the same behavior as
that of the superconducting sample. Our results suggest that strong
spin fluctuations remain even below T$_{SDW}$ in Co doped samples,
which may be an indication that Fe 3d spins of some part of the 3d
orbitals remain paramagnetic below T$_{SDW}$ as suggested by Singh
et al \cite{Singh, Athena1} based on Fermiology. For example, all 3d
spins are not ordered in e$_g$ but ordered in t$_{2g}$ orbitals, or
vise versa. In passing, we recall that $\frac{1}{T_1T}$ at
$^{139}$La sites in undoped LaFeAsO \cite{Ishida} and $^{75}$As
sites in undoped BaFe$_2$As$_2$ \cite{Takigawa, Iyo} is suppressed
by an order of magnitude or more below T$_{SDW}$.

The large in-plane resistivity below T$_{SDW}$ in Co-doped samples
\cite{Ahilan,Ning2} is probably related to these strong spin
fluctuations. It should also be noted that we find no signature of
additional spin freezing at low temperatures in either
$\frac{1}{T_1T}$ or $\Delta$B$_{hf}^c$. It is worth recalling that
in the case of lightly doped La$_{2-x}$Sr$_x$CuO$_2$ \cite{Chou,
Cho}, $\frac{1}{T_1T}$ at $^{139}$La sites shows additional
diverging behavior at T$_{sf}$, much below T$_N$. Furthermore,
B$_{hf}$ shows additional enhancement below T$_{sf}$. The spin
freezing temperature T$_{sf}$ turned out to be related to glassy
freezing of spin and charge stripes. Our present observation is
markedly different from the case of the lightly doped cuprates.

In conclusion, we have presented a $^{59}$Co and $^{75}$As NMR study
in the lightly electron doped, SDW ordered regime of
Ba(Fe$_{1-x}$Co$_x$)$_2$As$_2$. We demonstrated that Co doping
suppresses T$_{SDW}$, and changes the spin structure. The continuous
growth of the NMR linewidth below T$_{SDW}$ and the strong
enhancement of $\frac{1}{T_1T}$ at T$_{SDW}$ suggest a second order
phase transition into an SDW phase, most likely incommensurate with
the lattice and highly disordered. We did not detect any anomaly
from T$_{SDW}$ down to base temperature in either $\Delta$B$_{hf}^c$
or $\frac{1}{T_1T}$. This suggests the absence of freezing of
stripes or other analogous phenomena. On the other hand, large
$\frac{1}{T_1T}$ at T $\ll$ T$_{SDW}$ hints the residual
paramagnetic spins at $\textit{each}$ Fe site due to the
multi-orbital nature of FeAs layers. During the final stage of
preparing this manuscript, Bernhard et al. reported $\mu$SR
observation of static magnetism in a 4 $\%$ Co doped sample only
below 15 $\sim$ 20 K\cite{Bernhard}.

The work at McMaster was supported by NSERC, CIFAR and CFI. Research
sponsored by the Division of Materials Science and Engineering,
Office of Basic Sciences, Oak Ridge National Laboratory is managed
by UT-Battelle, LLC, for the U.S.
Department of Energy under contract No. DE-AC-05-00OR22725.\\


\end{document}